\newcommand{\ba}{\begin{array}}
\newcommand{\ea}{\end{array}}
\newcommand{\bean}{\begin{eqnarray*}}
\newcommand{\eean}{\end{eqnarray*}}
\newcommand{\rm}[1]{\mathrm{#1}}

\newcommand{\ep}{\epsilon}
\newcommand{\tr}{\mathrm{tr}}
\newcommand{\Tr}{\mathrm{Tr}}

\newcommand{\p}{\partial}

\newcommand{\bra}[1]{\left\langle {#1} \right|}
\newcommand{\ket}[1]{\left|  #1 \right\rangle}
\newcommand{\bracket}[3]{\left\langle {#1} \left| {#2} \right| {#3} \right\rangle \;}
\newcommand{\braket}[2]{\left\langle #1 | #2 \right\rangle \;}

\newcommand{\bmx}[1]{\left(\begin{array}{*{#1}{c}}}
\newcommand{\emx}{\end{array}\right)}
\newcommand{\bmxw}[1]{\renewcommand{\arraystretch}{2}\left(\begin{array}{*{#1}{c}}}
\newcommand{\bmxww}[1]{\renewcommand{\arraystretch}{2.5}\left(\begin{array}{*{#1}{c}}}
\newcommand{\bdet}[1]{\renewcommand{\arraystretch}{1.2}
	\left|\begin{array}{*{#1}{c}}}
\newcommand{\edet}{\end{array}\right|\renewcommand{\arraystretch}{1}}
\newcommand{\beq}{\begin{equation}}
\newcommand{\eeq}{\end{equation}}
\newcommand{\bea}{\begin{eqnarray}}
\newcommand{\eea}{\end{eqnarray}}

\newcommand{\ditem}[1]{\item[$\diamond$]}

\newcommand{\bit}{\begin{itemize}}
\newcommand{\eit}{\end{itemize}}

\newcommand{\eab}{\begin{eqnarray}}
\newcommand{\eae}{\end{eqnarray}}

\documentclass[aps,prb,twocolumn,superscriptaddress]{revtex4-1}
\usepackage{graphics}
\usepackage{epsfig}
\usepackage{amssymb}
\usepackage{color} 
\usepackage{subfigure}
\usepackage{ulem}

\begin{document}


\title{A Unified Formalism for calculating Polarization, Magnetization, and more in a Periodic Insulator}


\author{Kuang-Ting Chen}
\affiliation{Department of Physics, Massachusetts Institute of Technology, Cambridge, MA 02139}
\author{Patrick A. Lee}
\affiliation{Department of Physics, Massachusetts Institute of Technology, Cambridge, MA 02139}


\date{\today}

\begin{abstract}
In this paper, we propose a unified formalism, using Green's functions, to integrate out the electrons in an insulator under uniform electromagnetic fields. We derive a perturbative formula for the Green's function in the presence of uniform magnetic or electric fields. Applying the formula, we derive the formula for the polarization, the orbital magnetization, and the orbital magneto-polarizability, without assuming time reversal symmetry. Specifically, we realize that the terms linear in the electric field can only be expressed in terms of the Green's functions in one extra dimension. This observation directly leads to the result that the coefficient of the $\theta$ term in any dimensions is given by a Wess-Zumino-Witten-type term, integrated in the extended space, interpolating between the original physical Brillouin zone and a trivial system, with the group element replaced by the Green's function. This generalizes an earlier result for the case of time reversal invariance [see Z. Wang, X.-L. Qi, and S.-C. Zhang, Phys. Rev. Lett. {\bf 105}, 256803 (2010)].
\end{abstract}


\maketitle

\section{Introduction}

In recent years, there has been a series of development on how to calculate the  polarization and orbital magnetization quantum-mechanically in insulating systems, in terms of Bloch wave functions.\cite{niusemi,king,niuq,Thonhauser,souza,ceresoli} The polarization $\bf P$ measures the position differences between the band electrons and the lattice ions, which is in general nonzero in a crystal without inversion symmetry. In an open system, it results in boundary charges, which gives rise to the energy density $\Delta E={\bf -P\cdot E}$ in an external electric field. Inside the bulk, it is only defined modulo $ea$ where $e$ is the electron charge, and $a$ is the lattice spacing. This arises naturally from the ambiguity of associating an electron with a given ion. The reason why it has not been fully computed until rather recently is due to the unboundedness of the operator $\bf{r}$, which is the natural quantity to calculate the expectation value, when the lattice ions stay fixed. In Ref.~\cite{king}, they overcome this problem utilizing the Wannier orbitals. The Wannier orbitals are localized in an insulating system, and the matrix elements between them are well-defined, even if the operator itself is unbounded. They also provide an alternative method, which is to compute the current response under a change of the Hamiltonian without breaking the lattice translation symmetry. The integration of the current gives the change of the polarization. This method avoids calculating $\langle\textbf{r}\rangle$ altogether, at the expense of introducing an extra parameter and a Hamiltonian depending on it. This method also leaves the impression that only the diffrence of the polarization is properly defined. We, however, have argued otherwise in Ref.~\cite{mywork}. 

Similarly, the magnetization $\bf M$ is generally nonzero when the time-reversal symmetry is absent. The energy density change in the presence of an external magnetic field is given by $\Delta E=-{\bf M\cdot B}$. Usually the time-reversal symmetry is broken due to magnetism, and the dominant contribution to the magnetization is of spin origin. Nevertheless, via spin-orbit coupling the orbital motion contributes to the magnetization as well. The orbital magnetization is first computed in Ref.~\cite{niusemi,Thonhauser}, either by semi-classical methods\cite{niusemi}, or by calculating the matrix elements of ${\bf r\times v}$ using again the Wannier orbitals.\cite{Thonhauser} In the latter work, however, special care has to be taken toward the boundary, as they show that an extra term $M_{\rm IC}$ arising from the ``itinerent current" flowing around the boundary has to be included, in addition to the ``local current" contribution $M_{\rm LC}$, which involves the matrix element of ${\bf r\times v}$ between the bulk Wannier orbitals, to give the correct and the same answer as in the former calculation. Later on in Ref.~\cite{niuq}, the authors give a full quantum mechanical derivation, which calculates the energy of the system in a finite magnetic field based on the finite $q$ perturbation theory of the vector potential, and taking $q\rightarrow 0$ in the end. By taking a derivative with respect to $\bf B$ one gets the magnetization $\bf M$. 


After the discovery of the three-dimensional (3D) topological insulator\cite{hasankane}, it was shown in Ref.~\cite{QHZ} via a dimensional reduction procedure from one higher dimension, that the 3D topological insulator is characterized by the $\theta$-term, $\mathcal{L}_\theta\equiv\frac{\theta e^2}{4\pi^2}{\bf E\cdot B}$, with $\theta=\pi$ in the effective theory, where the electrons are integrated out. With time reversal symmetry, the coefficient $\theta$ is given by the integral of the Chern-Simons term of the Berry's connection in the Brillouin zone, which is independent of the gap size, has a $2\pi$ ambiguity, and therefore can only take the value $0$ or $\pi$. In Ref.~\cite{WXZ}, it is shown that with time reversal symmetry, the $\theta$ coefficient can also be written as a Wess-Zumino-Witten- (WZW-) type term, integrated in an extended space, where one interpolates between the system in question and a trivial insulator.

Since then, there have been various attempts to calculate this term explicitly in three spatial dimensions\cite{Mala,AAJD}, without assuming time reversal symmetry. The $\theta$-term is treated as an energy density, which can be understood as either the polarization in a magnetic field or the magnetization in an electric field, which are denoted as orbital magneto-polarizability (OMP) or orbital electro-susceptibility (OES). In Ref.~\cite{AAJD}, they offer two ways to calculate the OMP, similar to the methods described above for the polarization. In the first method they again calculate the current response to a change of Hamiltonian, but with a small uniform magnetic field turned on all the time. To perturb in the magnetic field they use the density matrix perturbation theory. In the second method, they evaluate the matrix element $\bf r$ between Wannier orbitals in the magnetic field, using the finite $q$ perturbation theory and then take the $q\rightarrow 0$ limit. In general, however, they find that $P^i=\alpha^i_jB^j$; that is, the OMP is not diagonal, and the $\theta$-term is just a part of it. This result is confirmed by Ref.~\cite{Mala}, where they use the Wannier orbitals to study the OES, by calculating the magnetization in an electric field. In this calculation, similar care has to be taken on the boundary. In either calculation, the diagonal part of $\alpha^i_j$ is defined modulo $\frac{e^2}{2\pi}$. The easiest way to understand this physically is the following. If we consider a cylinder geometry with the material in question in the bulk, an integer quantum Hall layer on the surface of the cylinder will change the diagonal response by $\frac{ne^2}{2\pi}$, where $n$ is the filling factor of the layer.

While all the results in the end agree with each other, the derivations are diverse, with various limitations and subtleties, as noted below:

 (i) The Wannier orbitals can only be defined when the Chern number of the bands is zero.\cite{thouless} Furthermore, to use the operator $\bf r$ or $\bf r\times v$, one is essentially limited to settings with open boundary conditions, as they are not well-defined on a torus. The boundary then has to be treated carefully, even if we are only interested in bulk properties: in Ref.~\cite{Thonhauser}, to get the correct expression for the magnetization, they have to consider two contributions, $M_{\rm LC}$ and $M_{\rm IC}$ as we briefly mentioned. The first term is the usual local matrix element between the bulk Wannier orbitals and the second term is itinerent, comes from the boundary, where the Wannier orbitals are deformed. The second term, however, can be written as a function of bulk parameters and is then argued to be present even in a setting with periodic boundary conditions. This is a subtle argument, because if we just start from the periodic system, there seems to be no reason to expect the second term. In fact, this argument reinforces that the matrix element $\bf{r\times v}$ cannot be used to represent the magnetization $\bf M$ in a setting with periodic boundary conditions, as the $M_{\rm IC}$ term will be missing.

(ii) The $q\rightarrow 0$ calculation is not justified in the first place. As we know, even for free electrons, the wave function in an uniform magnetic field forms Landau levels, which are not perturbatively connected to the plane waves, in arbitrarily small magnetic fields. This is due to the fact that the perturbation is expanding in powers of $\bf A(q)$ in stead of $\bf B(q)$, where $\bf B(q)=q\times A(q)$. If we do the perturbation formally anyway, in the limit $q\rightarrow 0$, but $\bf B$ remains finite, then the concern is that $\bf A(q)$ diverges. Indeed, in this setting we would find that the perturbed energy eigenfunctions are not orthogonal to one another at any given order. Even though the correct formulas are recovered when the $q\rightarrow 0$ limit is taken properly (probably due to the fact that we are actually calculating the physical properties at ${\bf B}=0$), it is certainly desirable to have a more reliable derivation.


(iii) The method of computing the response current to an adiabatic change of the Hamiltonian can be only applied to calculate the polarization. Magnetization change, for example, does not result in any bulk current flow, and thus cannot be computed in any similar way. Another potential problem is that this method only captures the change of polarization between the two systems. It is tempting, from the point of view of this method, to claim that only the difference of the polarization is physical. While one can always define the polarization of an atomic insulator to be zero and calculate between the interpolation of that and the state in question, it is not immediately obvious that any two state with different polarization are measurably different when they are separately put with periodic boundary conditions. A derivation without referring to any other Hamiltonian is therefore desirable, as this directly shows that, for example, the polarization is an intrinsic property, independent of boundary condtions. 

With the issues mentioned above in mind, we would therefore like to develop a formalism, which explains and computes everything mentioned in a unified manner. In addition, since the integer quantum Hall effect and its higher-dimensional analogs are closely related to the quantities mentioned above and can be derived using the Green's function techniques at finite momentum, we would like to propose a formalism utilizing Green's functions. In this paper, we provide such an unified formalism. In this formalism, we do not have to work with any boundary. We can also perturb in the uniform electromagnetic fields in a gauge-invariant way, without appealing to any finite momentum calculation. All the calculations are also done without changing the Hamiltonian. The paper is organized as follows: in the subsequent section, we introduce the formalism and outline the procedure. In Sec. III, we show the derivation of various quantities in detail.

\section{Formalism}
What are polarization and magnetization? With boundaries, they can be defined as charges and currents on the boundary; without boundaries, there has to be some inhomogeneity inside in order to observe the charges or currents. An alternative and more fundamental definition is that the polarization (magnetization) is the coefficient for the term proportional to $\bf E$($\bf B$) ,in the effective theory, when the electrons are integrated out. The boundary and the inhomogeneity charge or currents are then naturally derived when one solves the equation of motion of the effective theory, which is just the Maxwell equations in our case.

Therefore, our goal is to do the electronic part of the path integral, in the presence of the uniform electric field and the magnetic field as a back ground, perturbatively in $\bf E$ and $\bf B$. We then have ${\bf P}=-\p F/\p {\bf E}$, ${\bf M}=-\p F/\p {\bf B}$, and $\alpha^i_j=-\p^2 F/\p E^i\p B^j$, with $F=\beta^{-1}\log Z$ the free energy. This at first seems rather straight forward, as a standard diagrammatic procedure is readily available to calculate perturbative corrections to the partition function. Our goal seems no more than a one-loop calculation. It turns out not to be the case, however, when one looks carefully into the problem. The terms in the action of the effective theory we are after are total derivatives in terms of the electromagnetic gauge field. They are just zero in momentum space, where the standard procedure is carried out. This also reflects the difficulties mentioned in the introduction, as either the operator $\bf r$ or $\bf r\times v$ appear in the calculation of the polarization or the magnetization exactly due to the fact that $\bf E$ and $\bf B$ are spatial derivatives of the gauge potential.

To overcome this problem, we have to calculate in position space. Let us first deal with the magnetic field. While the wave function is not perturbative in the magnetic field, as we will show below, the gauge-invariant part of the Green's function is. The Green's function for a single-particle Hamiltonian satisfies the following equation:
\beq\label{simpleeq}
\sum_{x'}(\omega-H)_{xx'}g_{x'x''}=\delta_{xx''};
\eeq
$H$ is the single-particle Hamiltonian and $g$ is the Green's function of the electrons. Both are $n$ by $n$ matrices where $n$ labels the orbitals and spins. The system couples to a small uniform magnetic field via the Peierls substitution: 
\beq
H_{xx'}=(H_0+H')_{xx'}e^{\frac{ie}{\hbar}\int_x^{x'}\vec A\cdot dx},
\eeq
where $H_0$ is the Hamiltonian in zero field, $H'$ is some local perturbation that is proportional to $B$, e.g., the atomic diamagnetism, and $\vec A$ is the gauge potential. Since the correction to the Green's function as well as the free energy from $H'$ can be calculated in the standard way, we will set it to zero from now on. The line integral of the gauge potential follows a straight line from $x$ to $x'$.  In the following, we will use $A_{xx'}$ as the short-handed notation for $\int_x^{x'}\vec A\cdot dx$. We also set $e=\hbar=1$ when there is no ambiguity. Using the idea in Ref.~\cite{Finkel'stein}, this equation can be solved perturbatively in $B$ in the following way:

we write 
\beq\label{anz}
g_{xx'}=\tilde g_{xx'}e^{iA_{xx'}}
\eeq
 and notice that it does not change anything if we put $e^{iA_{xx''}}$ along with the $\delta$-function, we get
\beq
\sum_{x'}(\omega-H_0)_{xx'}e^{iA_{xx'}}\tilde g_{x'x''}e^{iA_{x'x''}}=\delta_{xx''}e^{iA_{xx''}}.
\eeq
Taking the exponential factor to the left-hand side, the three phases combine together, which gives the magnetic flux threading through the triangle formed by the three points $x, x', x''$. Independent of the gauge, we therefore have
\beq\label{solgreen}
\sum_{x'}(\omega-H_0)_{xx'}\tilde g_{x'x''}e^{iB\cdot(x'-x)\times(x''-x')/2}=\delta_{xx''}.
\eeq
Notice that this equation is now translationally invariant, and we can solve for $\tilde g$ to first order in $B$ by expanding the exponential and then Fourier transform, noting that $x$ can be replaced by $i\p/\p k$:
\beq
g_0^{-1}\tilde g-\frac {iB^c\ep^{abc}}2\frac{\p g_0^{-1}}{\p k^a}\frac{\p g_0}{\p k^b}=1;
\eeq
where $g_0\equiv g_0(k)=(\omega-H_0(k))^{-1}$; $H_0(k)=\sum_{x}H_{0,0x}\exp(-ikx)$ and $\tilde g$ is in Fourier space. We therefore get
\beq\label{main}
\tilde g=g_0+\frac {iB^c\ep^{abc}}2g_0\frac{\p g_0^{-1}}{\p k^a}\frac{\p g_0}{\p k^b}+\mathcal{O}(B^2).
\eeq
Notice that $\tilde g$ is gauge invariant. Once we have $\tilde g(k)$, the Green's function is just the inverse Fourier transform of it times the phase factor $e^{iA_{xx'}}$. We therefore have the real-space Green's function in the presence of the uniform magnetic field. 

While the calculation is straightforward, to our knowledge Eq.~(\ref{main}) is a new result. In Ref.~\cite{Finkel'stein}, without sources other than the magnetic field which breaks time reversal symmetry, this first order term vanishes and all they have to do is to set $\tilde g=g_0$. In that case, all the effect of the magnetic field comes from the phase. 

We can extend the calculation to include the perturbative correction in the uniform electric field as well. We start from the defining equation which is the Fourier transform of Eq.~(\ref{simpleeq}): 
\beq
(i\frac{\rm d}{{\rm d} t}-H)_{(x,x';t,t')}g(x',x'';t',t'')=\delta_{xx''}\delta(t-t'').
\eeq
Now we assume the coupling to the electric field comes from the space-time extension of the Peierls phase. Note that this procedure again does not include contributions from the response of the local orbitals to the electric field. We then use the same trick, define 
\beq\label{anz2}
g_{xx',tt'}=\tilde g_{xx',tt'}e^{iA_{xx',tt'}}
\eeq
where $A_{xx',tt'}$ is the line integral of the spacetime gauge field $(-V,\vec A)$ on the straight line connecting the two points. Following a similar procedure, noticing that $\frac{\p g_0^{-1}}{\p\omega}=1$, one can reach
\beq\label{maine}
\tilde g=g_0-\frac {iE^a}2\big(g_0\frac{\p g_0}{\p k^a}-\frac{\p g_0}{\p k^a}g_0\big)+\mathcal{O}(E^2).
\eeq
This procedure can easily be carried to arbitarary order of both the electric field and the magnetic field.

It is important to understand that Eq.~(\ref{anz}) and Eq.~(\ref{anz2}) are just a way to factor out the gauge dependence of the Green's function; it is not an approximation. The only approximation comes in when we Taylor-expand in powers of the flux threaded in the triangle formed by the three points.

Let us be concrete and give a specific example. Suppose we have a tight-binding system with $n$ orbitals sitting on each site. The $i$-th orbital is located at $\vec d_i$ from the lattice vector $\vec R$. Now the Hamiltonian $H_0$ is an $n\times n$ matrix in momentum space, and so is $\tilde g$. Notice that in deriving the formula, we have implicitly chosen the gauge such that $H_0(\vec k+\vec G)=U^\dag H_0(\vec k)U$; $U$ is a diagonal matrix with $U_{ii}=\exp(i\vec G\cdot \vec d_i))$. The boundary condition is similar for $\tilde g$. Our formula is then a matrix equation for the $n\times n$ matrix $\tilde g$. It is important that our formula only works with this ``twisted" boundary condition when there is a basis. 

We note that it has been shown earlier that the one-particle density matrix (OPDM) is also perturbative in the magnetic field, and can be calculated in a similar way.\cite{AAJD} Many quantities we calculate below can also be calculated using the OPDM. One key difference is that the Green's function can also be perturbed in powers of the uniform electric field as we have shown above. Combining the Berry's phase procedure as we will mention later on, the Green's functions formalism is thus a truly unified framework which can calculate perturbations of the uniform electromagnetic fields to arbitrary order, including the susceptibility and polarizability. The OPDM can always be derived from the Green's functions via  $\tilde\rho(k)=i\int\frac{{\rm d}\omega}{2\pi}\tilde g(k)$.

At zero temperature, without the electric field, the free energy is just the expectation value of $H$. The path integral can thus be performed by calculating the expectation value of the Hamiltonian in a uniform magnetic field. Other perturbations in the presence of the field can be captured in the usual way, replacing fermion bilinears with the Green's functions.

With a uniform electric field, the expectation value of the Hamiltonian is no longer the same as the free energy. We can understand this fact by taking the gauge $V=0$ (since our formulation is gauge independent.) In this gauge, the translational invariance in the time direction is lost, and one naturally does not expect any relation between the two quantities. One can directly see this by calculating the expectation value of the Hamiltonian in the presence of the electric field. We find that the expectation value of the Hamiltonian does not change with the electric field at first order.

How do we calculate the path integral in the presence of the electric field then? The following observation provides a hint: if we think about the imaginary-time path integral, the term $\bf{P\cdot E}$, unlike $\bf{M\cdot B}$, stays imaginary. Indeed as we discussed in Ref.~\cite{mywork}, the polarization $P^i$ is better thought of as a {\it Berry's phase}, instead of energy, when the gauge winding number in the $i$-direction is changed by one; i.e. 
\beq\label{berryp}
\phi_{\rm Berry}^i=-2\pi P_i.
\eeq
 Similarly, the extra Berry's phase in a magnetic field is related to the OMP by \beq
\Delta\phi_{\rm Berry}^i=-2\pi\alpha_{ij}\Phi_B^j,
\eeq
 where $\Phi_B$ is the total magnetic flux threading through the system. The 2$\pi$ ambiguity of both quantities thus comes naturally. For the sake of completeness, let us reiterate the procedure\cite{mywork,zak}:

Consider a system with periodic boundary conditions. To calculate accumulated the Berry's phase during a time when, for example, $\int A_x{\rm d}x$ increase by $2\pi$, first we shall consider how the single particle wave function changes as we increase $A_x$ uniformly. The Bloch wave function is given by
\beq
\psi_{nk}(x)=u_{nk}(x)e^{ikx}
\eeq
where $n$ is a band index; $u_{nk}(x)$ is periodic and satisfies
\beq
\left(\left(\nabla-(k+A_x)\right)^2+V(x)\right)u_{nk}(x)=E_{nk}u_{nk}(x).\label{uchange}
\eeq

As we increase $A_x$ uniformly to $A_x+\eta$, the momentum $k$ cannot change as it is fixed by the finite size $L$ and the periodic boundary condition. On the other hand, following Eq. (\ref{uchange}), $u_{nk}(x)$ changes as
\beq
u_{nk}(A_x+\eta)=u_{n(k-\eta)}(A_x)\label{uflow},
\eeq
which is just a corresponding shift of $k$ by $-\eta$. if $\eta=2\pi/L$, the system returns to its original state, but in a different gauge (i.e., with winding number different by one.) Notice that  while $u_{nk}(x)$ goes to the next avaiable value on the left, the $k$ in the exponential stays the same. The electronic wave function is therefore different from its starting state. Nevertheless, if we include the gauge field, the final state differs from the initial state by a large gauge transformation, and the Berry's phase accumulated in the process is well-defined.


Now we are ready to calculate the accumulated Berry's phase of the band electrons under the process, where the winding of the gauge field in the $x$-direction is increased by one. Let us define $\tilde A_x\equiv\int{\rm d}xA_x$:
\bea\label{berryd}
\phi_\mathrm{Berry}^x&=&i\int_{2\pi m}^{2\pi(m+1)}\rm d\tilde A_x\bra{\Psi_e}\frac{\partial}{\partial \tilde A_x}\ket{\Psi_e}\nonumber\\
&=&i\int_{2\pi m}^{2\pi(m+1)}{\rm d}\tilde A_x\sum_{k_i,n\in\rm{occ}}\bra{\psi_{nk_i}}\frac{\partial}{\partial \tilde A_x}\ket{\psi_{nk_i}}\nonumber\\
&=&i\sum_{k_i,n\in\rm{occ}}\int_{k_i}^{k_i+2\pi/L}{\rm d}k_x\bra{u_{nk}}\frac{\partial}{\partial k_x}\ket{u_{nk}}\nonumber\\
&=&i\int_{\rm BZ}{\rm d}k\sum_{n\in{\rm occ}}\bra{u_{n k}}\frac{\partial}{\partial k_x}\ket{u_{nk}}.
\eea
$\ket{\Psi_e}$ is the total electronic wave function; in the case we are interested it is just the Slater determinant of the occupied electron wave functions at the wave vectors $k_i$ allowed by the periodic boundary condition. In the second equality, we wrote the derivative acting on the Slater determinent as a sum of derivatives acting on single particle wave functions. In the third equality we then plug in the dependence of the wave functions, and change variables to $k$. Whenever $\tilde A_x$ increases by $2\pi$, each $u_{nk}$ reaches the next allowed eigenstate to the left by the periodic boundary condition (without actually changing the momentum eigenvalue.) As we sum over all the integral of eigenstates at different allowed $k$'s, the whole Brillouin zone (BZ) is covered exactly once and we reach the fourth equality. We can read out the expression of the polarization using Eq.~(\ref{berryp}):
\beq
P^x=-i\int_{\rm BZ}\frac{{\rm d}k}{2\pi}\sum_{n\in{\rm occ}}\bra{u_{n k}}\frac{\partial}{\partial k_x}\ket{u_{nk}}.
\eeq
This well-known result was derived in Ref.~\cite{king,zak}.

How do we express the Berry's phase in terms of the Green's functions? A naive thought would suggest that we cannot! Consider the ``gauge transform" defined by 
\beq\label{gauge}
\ket{u_{nk}}\rightarrow \exp(i\phi(k))\ket{u_{nk}}.
\eeq

We first observe that the Berry's phase, Eq.~(\ref{berryd}) is not invariant under this transform and may change by integer multiples of $2\pi$. The Green's function, on the other hand, is clearly invariant under this gauge transform. It is therefore impossible to express the Berry's phase solely in terms of polynomials of the Green's functions. 
\begin{figure}[htb]
	\centering
	\includegraphics[width=8cm]{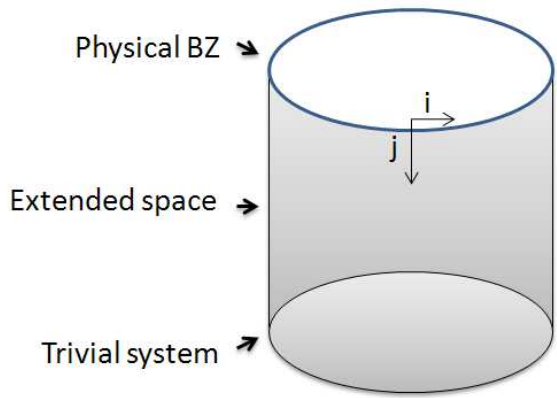}
	\caption{The black circle is the original physical space, with only the momentum direction along the electric field shown. The cylinder is the extended space, with the other end a trivial system. $i,j$ are orthonormal basis on the extended space;$i$ is along the direction of the eletric field, and $j$ points along the extra dimension.}
	\label{cartoon}
\end{figure} 
However, we then observe that the Berry's phase, expressed as a gauge-dependent loop integral in momentum space, can be cast as a gauge-invariant surface integral via the Stoke's theorem. We therefore extend our system to one extra dimension in the momentum space, interpolating between the original system and a trivial system whose Berry's phase is taken to be zero.\cite{WXZ,mynote} See Fig. 1 for an illustration. The gauge dependence, then, is hidden in the way we choose to extend the wave functions. The integrand on the surface can then be expressed in terms of the Green's functions, whose definition is also extended from the circle to the cylinder. The expression is as follows:
\bea\label{berry}
\phi_{\rm Berry}^i&=&i\oint_{\partial S}{\rm d}k^i\sum_{\alpha\in{\rm occ}}\bra{u_{k\alpha}}\frac{\partial}{\partial k^i}\ket{u_{k\alpha}}\nonumber\\
&=&i\int_S{\rm d}^2{\bf k}\epsilon^{ij}\partial_i\bra{\bf u_{k\alpha}}\partial_j\ket{\bf u_{k\alpha}}\nonumber\\
&=&\frac12\int_S{\rm d}^2{\bf k}\epsilon^{ij}\Tr'\left({\bf g}\p_i {\bf g}^{-1}\p_j {\bf g}\right);
\eea
$\p S$ is in the $i$-direction, which is the direction of the electric field, and $S$ is the enclosed cylinder in the extra dimension.  Bold quantities are extended into the enclosed surface. We require that the other boundary we extend to does not contribute to the Berry's phase. The trace with prime sums over all the bands (both occupied and unoccupied), and integrates over $\omega/2\pi$ as well as the other $(d-1)$ directions prependicular to $\bf E$. Please see Appendix A for a derivation.

 This construction naturally separates the integral into two contributions, as discussed below: for the expression to be dependent only on the boundary variables, the integrand on the surface has to be {\it closed}, in differential geometry terms. It is however, not always {\it exact}, in that integration over a closed surface does not always give you zero. A familiar example of an integrand being closed but not exact is $\nabla\theta$ on a circle, where $\theta$ is the polar angle. It is locally a total derivative; nevertheless when you integrate it over the entire circle it gives you $2\pi$. 

We can therefore separate the integral into two contributions. The first is {\it exact} and can be written as a total derivative in terms of Green's functions. It thus directly becomes a boundary integral via Stokes theorem. The remaining contribution cannot be written as a boundary integral in terms of the Green's functions. However, it has to be {\it topological}, meaning that it is invariant under smooth deformations which vanishes on the boundary. 
Topological integrands are of a specific form, as we will discuss in Appendix A. Specifically, in $d$ spatial dimensions the topological term is in the form of the Wess-Zumino-Witten (WZW) action\cite{witten} in $d+1$ spacetime:
\beq
I_{\rm wzw}^d=\int_{\p S} d^{d+2}x\ep^{a_1a_2...a_{d+2}}\prod_{i=1}^{d+2}(U\frac{\p}{\p x^{a_i}} U^{-1}),
\eeq
which is also defined in one extra dimensions. Our topological term will be of the same mathematical form, with the spacetime $x^{d+1}$ replaced by the physical momentum $k^d$ as well as the energy $\omega$, and the group element $U(x)$ replaced by the Green's function $g_0(k,\omega)$. Further more, the coefficient in front of the topological term is determined up to a sign by requiring that the value of the integral has a $2\pi$ ambiguity with different extension to the extended space. This point was made in Ref.~\cite{WXZ} for the case of time reversal invariance.

Let us summarize for the formalism: we have an expression for the Green's function in the presence of an uniform magnetic field. Without the electric field, we can calculate the expectation value of the Hamiltonian to get the logarithm of the partition function. We can also calculate the charge and current responses. If we want to capture the terms linear in the electric field, however, we have to calculate a Berry's phase, which can only be expressed in terms of the Green's functions in one extra dimension. In the following section we shall see detailed calculations for all the quantities mentioned above.

\section{Derivations}
In this section, we show in detail how we apply the formalism given in the previous section, to three different quantities: the charge response to the magnetic field, the magnetization, and the orbital magneto-polarizability (OMP). We also discuss the $\theta$-term, which is the isotropic part of the OMP, in higher dimensions.

\subsection{Charge response in a magnetic field.}
In an integer quantum Hall system in two dimensions, the magnetic field is locked with the density. This is also true for Chern insulators, which has a non-vanishing transverse conductivity in zero field. The transverse conductivity is related to the field derivative of the density by the Streda formula $\sigma_{xy}=\frac{\p \rho}{\p B}$. Here we first verify this result as a sanity check.

Starting from eqn. (\ref{main}), it is straight forward to calculate the charge density in the magnetic field:
\bea\label{charge}
\langle\rho\rangle&=&\frac{1}{L^2}\sum_{m,x}\langle c_{mx}^\dag c_{mx}\rangle=i\Tr(g)\nonumber\\
&=&\int \frac{{\rm d}^2k}{(2\pi)^2}\frac{{\rm d}\omega}{2\pi}i\;\tr\left( g_0+\frac {iB\ep^{ab}}2g_0\frac{\p g_0^{-1}}{\p k^a}\frac{\p g_0}{\p k^b}\right)\nonumber\\
&=&\rho_0+\frac{BC_{1}}{2\pi};
\eea
$C_1$ is the Thouless-Kohmoto-Nightingale-den Nijs (TKNN) index (Chern number) of the occupied bands. The capital trace in the first line implies summing over all the bands and integrating over all directions as well as energy with a facter of $(2\pi)^{-1}$ each. The lowercased trace in the second line implies summing all the bands only. The phase factor in eq. (\ref{anz}) is absent due to the fact that the creation operator and the annihilation operator are at the same position. The last equality is derived in Appendix A. Eq. (\ref{charge}) describes the locking of the density to the magnetic field. This ``incompressibility" is fundamental to quantum Hall physics and follows from charge conservation and Faraday's law when the magnetic field is adiabatically turned on. 

Nevertheless, it is somewhat intriguing to see this effect survive even on a torus with an uniform magnetic field, as our derivation implies. Without boundaries, the charge density can only change by adding or removing bulk states abruptly, even though the magnetic field is small and does not affect the energy gap in any appreciable way. In a tight-binding model, the states then must ``teleport'' between the occupied and empty bands. On a torus the magnetic flux has to be quantized; in the weak field limit, when we increase the magnetic flux by one flux quantum, there will be exactly $C_1$ states, "teleporting" from the unoccupied bands to the occupied bands, with out changing the energy gap in between. We have verified this phenomenon with a numerical diagonalization of an insulating system with $C_1\neq 0$.

This peculiarity becomes more apparent when one compare the findings with the usual linear response derivation. There the quantized conductivity or density change is derived from a bubble diagram at finite $q$, in the $q\rightarrow 0$ limit. The density modulates in the same way as the magnetic field, which becomes uniform only at the limit. There are no such teleportations of the states between the bands; the electrons flows from patches with a positive magnetic field to patches with a negative magnetic field (if the Chern number is positive) and vice versa. 

\subsection{Magnetization}
As sketched in the previous section, we calculate the orbital magnetization by computing the energy of the system in the presence of an uniform magnetic field, and the relation ${\bf M}=-{\rm d}\langle H\rangle/{\rm d}{\bf B}$. We continue our derivation in two dimensions, as the magnetization is inherently a two-dimensional phenomenon. We have
\bea
\langle H\rangle&=&\sum_{xx'}\int\frac{{\rm d}\omega}{2\pi} i\;\tr\left(H_{xx'}g_{x'x}\right)=\sum_{xx'}\int\frac{{\rm d}\omega}{2\pi}i\;\tr\left(H_{0xx'}\tilde g_{x'x}\right)\nonumber\\
&=&E_0-\frac {B\ep^{ab}}2\int\frac{{\rm d}^2k}{(2\pi)^2}\frac{{\rm d}\omega}{2\pi}\;\tr\left(H_0g_0\frac{\p g_0^{-1}}{\p k^a}\frac{\p g_0}{\p k^b}\right)
\eea
Plugging in $g_0=\sum_m\ket{u_m}\frac{1}{\omega-E_m}\bra{u_m}$ and noticing that the $\omega$-integral restricts the poles of the two $g_0$'s to be on opposite sides, we find that the derivative on $g_0$ can only act on the bra, and the derivative on $g_0^{-1}$ can act on the bra or ket but not the energy, in order for the expression not to vanish. We can simplify the term linear in $B$ to be
\bea
\langle\Delta H\rangle&=&-\frac{B\ep^{ab}}2\int\frac{{\rm d}^2k}{(2\pi)^2}\frac{{\rm d}\omega}{2\pi}\sum_{m,n}\theta(-E_mE_n)\\
&\times&\frac{E_m(E_n-E_m)}{(\omega-E_m)(\omega-E_n)}\braket{u_m}{\p_a u_n}\braket{\p_bu_n}{u_m}\nonumber\\
&=&\frac{iB\ep^{ab}}2\int\frac{{\rm d}^2k}{(2\pi)^2}\sum_{m,n}\theta(-E_mE_n)\nonumber\\
&\times&|E_m|\braket{u_m}{\p_a u_n}\braket{\p_bu_n}{u_m}.\nonumber
\eea
In the derivation we have used relations such as $\braket{\p_au_m}{u_n}=-\braket{u_m}{\p_au_n}$. Now we reexpress everything in terms of occupied bands only:
\bea\label{mag}
&&\sum_{m,n}\theta(-E_mE_n)|E_m|\braket{u_m}{\p_a u_n}\braket{\p_bu_n}{u_m}\nonumber\\
&=&\sum_{n\in {\rm occ}, m\in {\rm emp}}E_m\braket{u_m}{\p_a u_n}\braket{\p_bu_n}{u_m}\nonumber\\
&&-E_n\braket{u_n}{\p_a u_m}\braket{\p_bu_m}{u_n}\nonumber\\
&=&\sum_{n,n'\in {\rm occ}}\bracket{\p_bu_n}{H}{\p_au_n}-E_{n'}\braket{u_{n'}}{\p_a u_n}\braket{\p_bu_n}{u_{n'}}\nonumber\\
&&\nonumber\\
&&-E_n\braket{\p_au_n}{\p_bu_n}+E_n\braket{\p_au_n}{ u_{n'}}\braket{u_{n'}}{\p_bu_n}\nonumber\\
&=&\sum_{n\in {\rm occ}}\bracket{\p_bu_n}{H}{\p_au_n}-E_n\braket{\p_au_n}{\p_bu_n}.
\eea
Putting eq. (\ref{mag}) back into the expression for the energy and dividing it by $B$, we recover the result derived in Ref.~\cite{niuq,Thonhauser}:
\beq\label{magnetization}
M=\frac{i\ep^{ab}}2\int\frac{{\rm d}^2k}{(2\pi)^2}\sum_{n\in {\rm occ}}\bracket{\p_au_n}{H+E_n}{\p_bu_n}
\eeq
 Specifically, the first term is called $M_{\rm LC}$ and the second term is called $M_{\rm IC}$ in Ref.~\cite{Thonhauser}. One can also derive the same expression using the OPDM.\cite{gonze}.

From our derivation, both terms come together from one simple expression in the bulk. If we have a boundary, the current flowing on the edge is the sum of the two, as is required by the Maxwell equations, ${\bf J}=\nabla\times{\bf M}$. One might be tempted to conclude that the two terms generate separately measurable currents from the derivation in Ref.~\cite{Thonhauser}, but that is not the case. One can, however, decompose it into two contributions (different from but related to $M_{\rm{LC}}$ and $M_{\rm{IC}}$) where the difference between the two terms can also be measured.\cite{souza}

In addition, as also discussed in Ref.~\cite{ceresoli}, eq. (\ref{magnetization}) naturally shows that there are $C_1$ gapless edge states for a Chern insulator with Chern number $C_1$ in two spatial dimensions. Suppose we shift the chemical potential by $\Delta\mu$. Now the magnetization changes by $C_1\Delta\mu/2\pi$, which implies the current on the edge changes by the same amount. If we think from the edge perspective, a chemical potential change of $\Delta\mu$ implies that the density of occupied edge states increases by $C_1\Delta\mu/2\pi v$, and carries additional current $C_1\Delta\mu/2\pi$. The two observations would not have matched, had we only taken the contribution $M_{IC}$ as the current flowing around the edge.

On the other hand, we have the same orbital magnetization formula for a system without boundary. This means that if the system is a Chern insulator, its magnetization will depend on the chemical potential, even if there are no edges. This somewhat puzzling observation actually comes from the density locking to the magnetic field mentioned in the previous subsection. Since we define the magnetization as the derivative of the free energy with respect to $B$, with $F=E-\mu N$, the free energy change with the magnetic field does depend on $\mu$. It is, however, unclear to us whether the``magnetization" defined this way is a measurable quantity on a periodic system.

With the discussions above, we therefore conclude that the magnetization of an insulator is a bulk quantity. The current flowing on the edge can always be derived via the equation of motion of the effective theory, i.e., the Maxwell equations. We also note that the orbital magnetic susceptibility can be calculated by expanding the Green's function to second order in $B$.

\subsection{Electric Polarizability}
The calculation of the polarization is already covered in the previous section and in Ref.\cite{mywork}. While we can express it in terms of the Green's function in one extra dimension as done in Eq. (\ref{berry}), after integrating out $\omega$ and integrating back to the boundary, the result is just what we start with. As a nontrivial example of using the formalism, here we calculate the polarizability, by considering the first order correction of the Green's function in a uniform electric field. We start from Eq.~(\ref{berry}), plug in Eq.~(\ref{maine}), and notice that there is an additional phase proportional to the electric field from contracting the three green's functions in real spacetime. (Notice that the time-dependent gauge potential does not break translational symmetry in the spatial direction so Eq.~(\ref{berry}) still applies. The trace in the time direction, however, has to be carried out in real space; it is easier just to imagine every thing is done in real space and then converted back.) We thus have
\bea
\Delta\phi^i_{\rm Berry}&=&I_1+I_2;\nonumber\\
I_1&=&\frac {iE^a}4\int_S{\rm d}^2{\bf k}\ep^{ij}\Tr'\bigg(({\bf g}_0\p_a{\bf g}_0-\p_a{\bf g}_0{\bf g}_0)\p_i{\bf g}_0^{-1}\p_j{\bf g}_0\nonumber\\
&+&{\bf g}_0\p_i{\bf g}_0^{-1}\p_j({\bf g}_0\p_a {\bf g}_0-\p_a{\bf g}_0{\bf g}_0)\bigg)\nonumber\\
I_2&=&-\frac{iE}{4}\int_S{\rm d}^2{\bf k}\ep^{ij}\Tr'\bigg(\p_\omega{\bf g}_0\p_a\p_i{\bf g}_0^{-1}\p_j{\bf g}_0\bigg).
\eea
Taking advantage of the relation $\p_\omega{\bf g}_0^{-1}=1$ and $\p_\omega{\bf g}_0=-{\bf g}_0^2$, we can simplify the expression to
\bea
\Delta\phi^i_{\rm Berry}&=&-\frac {iE^a}4\int_S{\rm d}^2{\bf k}\ep^{ij}\Tr'\bigg(\p_a(\p_\omega {\bf g}_0\p_i{\bf g}_0\p_j{\bf g}_0)\nonumber\\
&+&2\p_i({\bf g}_0\p_a\p_j{\bf g}_0)\bigg)\nonumber\\
&=&-i\pi E^a\int\frac{{\rm d}\omega}{2\pi}\int_{BZ}\frac{{\rm d}^dk}{(2\pi)^d}\;\tr\big(g_0\p_a\p_i g_0\big).
\eea
Notice the only total derivative that does not vanish after integration has to be along the extra dimension. As we see here, there is no topological contribution in the polarizability; the integral thus has to reduce to a total derivative, and we can just integrate it back to the physical Brillouin zone. Using Eq.~(\ref{berryp}), we get the polarizability tensor
\beq
\ep_{ij}=\frac i2\int\frac{{\rm d}\omega}{2\pi}\int_{BZ}\frac{{\rm d}^dk}{(2\pi)^d}\;\tr\big(g_0\p_i\p_j g_0\big),
\eeq
which is the same expression as one would get using the usual perturbation theory at finite momentum $q$, and take the $q\rightarrow 0$ limit.

\subsection{Orbital Magneto-Polarizability}
We can calculate the OMP by considering the polarization in a uniform magnetic field in three spatial dimensions. We start from Eq.~(\ref{berry}) and plug in the Green's function in the presence of the magnetic field. Recall that $i$ is the direction of the electric field, and $j$ is the direction of the extra dimension:
\bea\label{Berry3}
\phi_{\rm Berry}^i&=&\frac12\int_S{\rm d}^2{\bf k}\ep^{ij}\Tr'\left({\bf g}\p_i{\bf g}^{-1}\p_j {\bf g}\right)\nonumber\\
&\equiv&2\pi^2\ep^{ij}L^2\Tr^S\left({\bf g}\p_i{\bf g}^{-1}\p_j {\bf g}\right)
\eea
Here $\bf g$ denotes the Green's function at a given momentum in the $(i,j)$ surface. It is in general not translationally invariant in the remaining directions. Here we introduce the new notation $\Tr^S$ for later convenience. $\Tr^S$ is defined as integrating over momentum divided by $(2\pi)^2$ in the $(i,j)$ direction within the boundary, integrating over $\omega/2\pi$, and summing over positions, divided by $L$ in the remaining directions if they are not translationally invariant; the lattice is replaced by integration over the momentum and divided by $(2\pi)$ otherwise. When $\bf E$ and $\bf B$ are not prependicular, we have to take a Landau gauge to make the Green's function translationally invariant in the direction of the electric field for the derivation in the previous section to work; it nevertheless does not affect the result, Eq.~(\ref{Berry3}), and our calculation below.

Now we plug in Eq.~(\ref{main}). To first order in $\bf B$, not only do we have $\bf \tilde g$ to first order, but we also have to consider that the three product of $\bf g$'s, contains three phases which sums to be the flux threading through the triangle. Indentical to what we did in Sec. II, we Fourier transform, Taylor-expanding the phase to first order in $\bf B$. The result is

\bea\label{omp}
\Delta\phi_{\rm Berry}^i&=&I_1+I_2;\nonumber\\
I_1&=&\pi^2iB^cL^2\ep^{abc}\ep^{ij}\Tr^S\big({\bf g}_0\p_a{\bf g}_0^{-1}\p_b{\bf g}_0\p_i{\bf g}_0^{-1}\p_j{\bf g}_0\nonumber\\
&+&{\bf g}_0\p_i{\bf g}_0^{-1}\p_j({\bf g}_0\p_a{\bf g}_0^{-1}\p_b{\bf g}_0)\big)\nonumber\\
I_2&=&-\pi^2iB^cL^2\ep^{abc}\ep^{ij}\Tr^S\left(\p_a {\bf g}_0\p_b\p_i {\bf g}_0^{-1} \p_j {\bf g}_0\right);\nonumber\\
\eea 
$I_1$ is from first-order terms in $\bf \tilde g$ and $I_2$ is from the phase. $\bf g_0$ is the Green's function in zero magnetic field in the extended space. The trace now indicates integration in all the momentum directions as well as $\omega$ and divided by $(2\pi)$ in each direction, as the translational invariance is restored. $I_1$ can be rewritten as
\bea\label{first}
I_1&=&\pi^2iB^cL^2\ep^{abc}\ep^{ij}\Tr^S\big({\bf g}_0\p_a{\bf g}_0^{-1}\p_b{\bf g}_0\p_i{\bf g}_0^{-1}\p_j{\bf g}_0\nonumber\\
&+&(i\leftrightarrow a, j\leftrightarrow b)
-\p_j(\p_i{\bf g}_0\p_a{\bf g}_0^{-1}\p_b{\bf g}_0)\big)
\eea
in which the last term in the second line can readily be integrated back to the physical momentum space. The first two terms are almost in the form of the topological terms we mentioned in Appendix A, but without complete antisymmetrization among the indices.

Let us now look at $I_2$. We would like to separate this term into a total-derivative and some remaining parts which give the topological term; to achieve this, we need to take advantage of the condition  $\p_\omega {\bf g}_0^{-1}=1.$ Note that this relation also implies $\p_\omega\p_a {\bf g}_0^{-1}=0$ and $\p_\omega {\bf g}_0=-{\bf g}_0^2.$ By inserting $\p_\omega {\bf g}_0^{-1}$ at the end of the term and integrating the $\omega$-integral by parts, we find (from here on we omit the subscript of ${\bf g}_0$ to avoid cluttering in the equations):
\bea
0&=&\Tr^S\bigg({\bf g}\p_a {\bf g}^{-1}{\bf g}(\p_b\p_i {\bf g}^{-1}){\bf g}\p_j {\bf g}^{-1}{\bf g}\nonumber\\
&+&{\bf g}(\p_b\p_i {\bf g}^{-1}){\bf g}\p_j {\bf g}^{-1}{\bf g}\p_a {\bf g}^{-1}{\bf g}\nonumber\\
&+&{\bf g}\p_j {\bf g}^{-1}{\bf g}\p_a {\bf g}^{-1}{\bf g}(\p_b\p_i {\bf g}^{-1}){\bf g}\bigg).
\eea
For convenience, we use the notation $(\bf abij)$ to stand for $\Tr^S({\bf g}\p_a{\bf g}^{-1}{\bf g}\p_b{\bf g}^{-1}{\bf g}\p_i{\bf g}^{-1}{\bf g}\p_j{\bf g}^{-1}{\bf g})$, $\bf (a[bi]j)$ to stand for $\Tr^S({\bf g}\p_a {\bf g}^{-1}{\bf g}(\p_b\p_i {\bf g}^{-1}){\bf g}\p_j {\bf g}^{-1}{\bf g})$, ..., etc. The equation above becomes
\beq{\bf
(a[bi]j)+([bi]ja)+(ja[bi])}=0.
\eeq
Also, Eq. (\ref{first}) becomes
\beq\bf 
(abij)+(ijab)-\p_j(iab);
\eeq
We then integrate by parts twice on the second and the third term, noticing that $a,b$ and $i,j$ are separately antisymmetrized, to make them into the form of the first:
\bea\bf
([bi]ja)&=&\bf (a[bi]j)+(ibja)-(bjia)+\p_i(bja);\\
\bf (ja[bi])&=&\bf (a[bi]j)+(jaib)-(jbai)-\p_j(bai).
\eea
We therefore have
\bea\bf
-(a[bi]j)&=&\frac13\big({\bf -(bjia)-(jbai)+(ibja)+(jaib)}\nonumber\\
&+&\bf \p_i(bja)-\p_j(bai)\big).
\eea
Now we can sum over all contributions, and get
\bea
\Delta\phi_{\rm Berry}^i&=&\pi^2iB^cL^2\ep^{abc}\ep^{ij}\bigg(\bf (abij)+(ijab)\nonumber\\
&+&\frac13\big({\bf (bija)+(jabi)+(ibja)+(jaib)}\big)\nonumber\\
&-&{\bf\p_j(iab)}+\frac13\big({\bf\p_i(bja)-\p_j(bai)}\big)\bigg)
\eea 
Notice that the integrating-by-part trick in the $\omega$-direction can also be applied to expressions such as $\bf(abij)$ and $\bf(iab)$, and similarly we get
\bea
{\bf(abij)+(bija)+(ijab)+(jabi)}=0\\
{\bf(aibj)+(ibja)}=0\\
{\bf(iab)+(abi)+(bia)}=0\label{26}
\eea
Therefore, by writing
\beq
{\bf(abij)+(ijab)}=\frac13\left({\bf(abij)+(ijab))}\right)-\frac23\left({\bf(bija)+(jabi)}\right),
\eeq
the "topological part" of $\phi_{\rm Berry}$ is
\bea
\Delta\phi_{\rm Berry,wzw}^i=\frac13\pi^2iB^cL^2\ep^{abc}\ep^{ij}\bigg({\bf(abij)+(ijab)}\nonumber\\
{\bf+(aijb)+(iabj)+(ibja)+(ajbi)}\bigg).
\eea
Notice that this term is totally antisymmetric in all the indices. This is expected, as it is topological only when all the indices are antisymmetrized. It is also of the form of the WZW action if we put $\p_\omega g^{-1}$ into the expression.

Since the direction of the magnetic field is prependicular to the $(ab)$ plane, it has to be in the $(ij)$ plane, for the topological term not to vanish. This implies that only the component of the magnetic field in the direction of the electric field contributes in the topological part. Gathering everything we finally have
\beq
\Delta\phi_{\rm Berry, wzw}^i=\frac{\pi^2i}3 \Phi_B^i\ep^{abcd}{\bf(abcd)}
\eeq
The remaining part can be reorganized using Eq.~(\ref{26}):
\beq
\Delta\phi_{\rm Berry, 3d}^i=-\frac{2\pi^2i}{3}\Phi_B^c\ep^{abc}\ep^{ij}\big({\bf\p_j(bai)+\p_j(iab)}\big)
\eeq
Integrating back to the physical momentum space, we have
\beq
\Delta\phi_{\rm Berry,3d}^i=\frac{\pi i}3\Phi_B^c\ep^{abc}\big((i ab)+(bai)\big).
\eeq
Here $(iab)$ stands for $\Tr\big(g\p_ig^{-1}g\p_ag^{-1}g\p_bg^{-1}g\big)$ with the trace summing over the energy as well as the physical momentum directions, with $(2\pi)^{-1}$ in every direction. The difference of a factor of $(2\pi)$ in front comes from the different number of $(2\pi)^{-1}$ in the definition of the traces $\Tr^S$ and $\Tr$, in four and three spatial dimensions respectively. Combining, we thus have our final answer in terms of the Green's functions:
\bea\label{ompg}
\alpha_{ij}&=&(\alpha_{\rm wzw}+\alpha_{\rm 3d})_{ij},\nonumber\\
{\alpha_{\rm wzw}}_{ij}&=&-\frac{\pi i}6 \ep_{abcd}\Tr^S({\bf g}\p_a{\bf g}^{-1}{\bf g}\p_b{\bf g}^{-1}{\bf g}\p_c{\bf g}^{-1}{\bf g}\p_d{\bf g}^{-1}{\bf g})\delta_{ij};\nonumber\\
{\alpha_{\rm 3d}}_{ij}&=&-\frac{i}6\ep_{abj}\Tr\big(g\p_ig^{-1}g\p_ag^{-1}g\p_bg^{-1}g-h.c.\big).
\eea
Notice that in terms of the Green's functions, $\alpha_{\rm wzw}$ can only be expressed with the extended dimension. Eq. (\ref{ompg}) generalizes the result in Ref.~\cite{WXZ} to the generic time-reversal breaking cases, where the WZW integral can take continuous values. Note that there is an additional term $\alpha_{\rm 3d}$ which is zero in the time reversal invariant case.

To get the expression entirely in terms of variables in the physical momentum space, we have to expand the Green's functions explicitly in the eigenbasis then integrate it back to the physical momentum space. Taking advantage of the topological property of $\alpha_{\rm wzw}$ and using Eq.~(\ref{ratio}) in Appendix A, we can immediately know that the first term contains a part which can be expressed using the Berry's phase gauge field strength, and some other part which is a global total derivative, and invariant under the gauge transform defined by Eq.~(\ref{gauge}). When integrated back to the physical momentum space, the first part becomes the Chern-Simons term with the Berry's phase gauge field $\mathcal{A}_{\mu,nn'}\equiv \bra{u_{nk}}-i\frac{\partial}{\partial k^\mu}\ket{u_{n'k}}$; the remaining part combined with $\alpha_{\rm 3d}$ gives the rest of the tensor $\alpha_{ij}$ as derived in Ref.~\cite{AAJD}:
\bea\label{omp2}
\alpha_{ij}&=&(\alpha_{\rm CS}+\alpha_{\rm G})_{ij};\nonumber\\
{\alpha_{\rm CS}}_{ij}&=&-\frac12\delta_{ij}\int\frac{{\rm d}^3k}{(2\pi)^3}\epsilon^{abc}\tr(\mathcal{A}_a\partial_b\mathcal{A}_c+i\frac23\mathcal{A}_a
\mathcal{A}_b\mathcal{A}_c);\nonumber\\
{\alpha_G}_{ij}&=&\frac{1}2\ep_{abj}\int\frac{{\rm d}^3k}{(2\pi)^3}\sum_{m\in{\rm emp},n\in{\rm occ}}\bigg(\nonumber\\
&&\frac{\braket{\p_in}{m}\bracket{m}{\{\p_a H,\p_b \mathcal{P}\}}{n}}{E_n-E_m}+c.c.\bigg),
\eea
where $\mathcal{P}\equiv\sum_{n\in{\rm occ}}\ket{n}\bra{n}$ is the projector to the occupied bands. The detail of the calculation is shown in Appendix B.

\subsection{$\theta$-term in higher dimensions}
While it is straight forward to generalize the complete calculation in the previous section to obtain all components of the anolog of OMP in higher dimensions, the totally-antisymmetric part, i.e., the $\theta$-term, is especially easy to compute. Here as an illustration, we calculate the coefficient $\theta_{\rm 5d}$ of the $\theta$-term in (5+1)D as defined below:
\beq\label{deft}
\mathcal{L}_\theta=\frac{\theta_{\rm 5d}}{384\pi^3}\ep^{\alpha\beta\gamma\delta\mu\nu}
F_{\alpha\beta}F_{\gamma\delta}F_{\mu\nu}.
\eeq
First we find the Green's function to second order of the magnetic field. In higher dimensions, the term in the exponential in Eq. (\ref{solgreen}) becomes $[iF_{ab}(x'-x)^a(x''-x)^b/2]$. Since we are only interested in contributions with all the indices antisymmetrized, only first derivatives will contribute. When we Taylor-expand Eq. (\ref{solgreen}) to second order and Fourier transform, we get
\bea
\tilde g_2&=&-\frac14(F_{ab}F_{cd}+F_{ad}F_{bc}+F_{ac}F_{db})\\
&&g_0(\p_ag_0^{-1})g_0(\p_bg_0^{-1})g_0(\p_cg_0^{-1})g_0(\p_dg_0^{-1})g_0+\dots,\nonumber
\eea 
where $\tilde g_2$ is the second order term of $\tilde g$, and the indices run through all five spatial directions. The $(\dots)$ vanishes when we antisymmetrize all the indices. Plugging into Eq. (\ref{berry}), we then have
\bea
\phi_{\rm Berry}^i&=& -\frac{3\pi^2}2\ep^{ij}L^4(F_{ab}F_{cd}+F_{ad}F_{bc}+F_{ac}F_{db})\nonumber\\
&&{\bf(ijabcd)}+\dots,
\eea
where we have used the abbreviated notation introduced in the previous section. We still only need to keep track of the parts which do not vanish after antisymmetrizing all the indices. Antisymmetrizing, noticing that in each direction (say $E_x$, $B_{yz}$, $B_{uv}$) summing over indices gives a factor of 8, we then have
\beq
\phi_{\rm Berry}^{\rm iso}=-\frac{\pi^2}{60}\ep^{abcdef}\big({\bf abcdef}\big)\Phi^1_B\Phi^2_B,
\eeq
$\Phi^1_B$ and $\Phi^2_B$ are the two magnetic fluxes threading through the four directions prependicular to the electric field. Notice that from the definition Eq. (\ref{deft})$, \theta_{\rm 5d}$ is  exactly the Berry's phase when the flux threading through each direction equals $2\pi$, we then have
\bea\label{t5d}
\theta_{\rm 5d}&=&-\frac{\pi^4}{15}\ep^{abcdef}\big({\bf abcdef}\big)\nonumber\\
&=&-\frac{1}{1920\pi^3}I^6_F;
\eea 
$I^6_F$ is defined in Eq. (\ref{i2df}). This is the higher-dimensional analog of the trace of $\alpha_{\rm wzw}$, which includes both the second Chern-Simons term $\mathcal{A\wedge F\wedge F}$, and some other inter-gap contributions. To reexpress $\theta_{\rm 5d}$ entirely in terms of Bloch wave functions and energies in the physical Brillouin zone, however, is rather tedious, and we shall not do it here.

\section{Discussion}
In this paper, we provide a formalism to integrate out the electrons with external uniform electromagnetic fields. This formalism provides a unified and systematic way to calculate quantum-Hall type responses, polarization, polarizability, orbital magnetization, orbital magnetic susceptibility, and OMP. From the perspective of the formalism, all of the quantities mentioned are of bulk nature, and all calculations can be done with periodic boundary conditions. The existence of the edge current or response can be derived from the equation of motion of the resulting effective theory, which is defined in the bulk, independent of the boundary conditions.

In our formulation, one key insight is that the linear term in the magnetic field is an energy, whereas the linear term in the electric field is a Berry's phase. This explains why the polarization and the OMP are defined only modulo $2\pi$ in certain units, whereas the magnetization is always rigorously defined. 

The wave function under an uniform magnetic field is nonperturbative; nevertheless, the gauge invariant part of the Green's function is perturbative, even strictly at $q=0$. In Eq. (\ref{main}), the only expansion parameter is the flux enclosed in the triangle; inside an insulator with finite range correlation functions, the expansion is controlled. 

When the $\theta$-term is first discovered in the 3D topological insulators, it was shown for time-reversal invariant systems, $\theta$ is given by the Chern-Simons term of the Berry's phase gauge field defined in momentum space. It is not until much later that extra contributions which depend on the inter-gap matrix elements as well as the gap size are discovered for the general case without time reversal invariance, along with the off-diagonal components. From our calculation, two terms come together naturally, and are most simply expressed as an integral of the WZW term with Green's functions in the extended space. The same conclusion holds for the higher dimensional $\theta$ term as well. 

However, we have to note that it is not easy to convert the expression in the extended space back to the Bloch wave functions and energies in the physical Brillouin zone. For this purpose, the density matrix perturbation theory formalism\cite{mywork} seems to be more useful. Nevertheless, from those methods it is harder to obtain the Chern-Simons term; it is also not as straight-forward to generalize to higher dimensions. Our formula Eq. (\ref{ompg}) and the higher dimensional generalization Eq. (\ref{t5d}) thus complement the other methods and offer a better conceptual understanding.

There is an important difference between the derivation of the $\theta$-term using our approach and the dimensional reduction procedure used in Refs.~\cite{QHZ,WXZ}. Even though they give the same result with time reversal symmetry, the latter is not readily generalizable to the general case without time reversal symmetry. In the dimensional reduction procedure, one thinks of the system in one higher dimension as a collection of systems in the physical dimensions with different $\theta$. The sum of the $\theta$-term of the systems in the physical dimensions gives rise to the quantized transverse response in one higher dimension:
\beq
\sigma_{2d}\sim\oint \frac{{\rm d}k}{2\pi}\frac{\p\theta_{2d-1}}{\p k},
\eeq
$\sigma_{2d}$ is the coefficient of the quantized transverse response in $2d$ spatial dimensions, and $k$ is along the direction of the extra dimension. For $\sigma_{2d}$ to be nonzero,  $\theta_{2d-1}$ has to be gauge dependent with the gauge transform defined in Eq. (\ref{gauge}). From the expression of $\sigma_{2d}$, all one can deduce is that $\theta_{2d-1}$ must contain the $(2d-1)$-dimensional analog of either $\alpha_{\rm wzw}$ or $\alpha_{\rm CS}$, as well as some arbitrary quantity which is gauge invariant. In three spatial dimensions, $\theta$ is odd under time reversal symmetry; all the gauge invariant terms then are required to be zero when the symmetry is present. This is the only situation in which $\alpha_{\rm wzw}$ and $\alpha_{\rm CS}$, given in Eq. (\ref{ompg}) and Eq. (\ref{omp2}) respectively, are equal, and the derivation in Ref.~\cite{WXZ,QHZ} unambiguously determines the answer.

\section*{Acknowledgement}
We thank Karen Michaeli for useful discussions. The work is under the support of NSF grant DMR 1104498.

\section*{Appendix A: Topological combination of Green's functions}
In this section, we first discuss what kind of combinations of Green's functions are topological, then we provide some formulas we have used in the main text.

When we say an integral is topological, we mean that the value of the integral is independent of any smooth deformations that leaves the integrand on the boundary unaltered. For example, the following integral is topological:
\bea
I&=&\int{\rm d}\omega\; \tr\big(g\p_\omega g^{-1}\big)\\
\delta I&=&\int{\rm d}\omega\;\tr\big(\delta g\p_\omega g^{-1}+g\p_\omega \delta g^{-1}\big)\nonumber\\
&=&\int{\rm d}\omega\;\tr\big(\delta g\p_\omega g^{-1}+\p_\omega(g \delta g^{-1})-\p_\omega g \delta g^{-1}\big)\nonumber\\
&=&\int{\rm d}\omega\;\tr\big(\delta g\p_\omega g^{-1}+\p_\omega(g \delta g^{-1})+\p_\omega g g^{-1}\delta gg^{-1}\big)\nonumber\\
&=&\int{\rm d}\omega\;\tr\big(\p_\omega(g \delta g^{-1})\big);
\eea
we have used $\delta (gg^{-1})=\p_\omega(gg^{-1})=0$ in the last two equailties.
We see that the variation $\delta I$ is a total derivative of a single-valued function, which implies that it is zero if the integrand is not varied on the boundary. Notice here we did not assume any particular form of $g$; that is, $I$ is topological under arbitrary smooth deformation of $g$.

We can construct similar topologically-invariant integrals in higher dimensions. In fact, 
\beq
I^{2d}=\int{\rm d}\omega{\rm d}^{2d}k\;\ep^{a_1a_2...a_{2d+1}}\tr\big(\prod _{i=1}^{2d+1}(g\p_{a_i}g^{-1}) \big)
\eeq
is topological. To prove this, we notice first that when we have an even number of $(g\p_{a_i}g^{-1})$ multiplied together with their indices antisymmetrized, they become a total derivative:
\bea 
&&\ep^{a_1a_2...a_{2d}}\prod _{i=1}^{2d}(g\p_{a_i}g^{-1})\nonumber\\
&=&\ep^{a_1a_2...a_{2d}}(-1)^d\prod_{j=1}^d(\p_{a_{2j-1}}g)(\p_{a_{2j}}g^{-1})\\
&=&\ep^{a_1a_2...a_{2d}}(-1)^d\p_{a_1}\big(g\p_{a_2}g^{-1}\prod_{j=2}^d(\p_{a_{2j-1}}g)(\p_{a_{2j}}g^{-1})\big)\nonumber.
\eea
Now we consider a general deformation:
\begin{widetext}
\bea
\delta I^{2d}&=&(2d+1)\int{\rm d}\omega{\rm d}^{2d}k\;\ep^{a_1a_2...a_{2d+1}}\tr\big(\delta(g\p_{a_1}g^{-1})\prod _{i=2}^{2d+1}(g\p_{a_i}g^{-1}) \big)\nonumber\\
&=&(2d+1)\int{\rm d}\omega{\rm d}^{2d}k\;\ep^{a_1a_2...a_{2d+1}}\tr\big((\delta g\p_{a_1} g^{-1}+\p_{a_1}(g \delta g^{-1})+\p_{a_1} g g^{-1}\delta gg^{-1})\prod _{i=2}^{2d+1}(g\p_{a_i}g^{-1}) \big)\nonumber\\
&=&(2d+1)\int{\rm d}\omega{\rm d}^{2d}k\;\ep^{a_1a_2...a_{2d+1}}\tr\big((\delta g\p_{a_1} g^{-1}-g\p_{a_1}g^{-1}\delta gg^{-1})\prod _{i=2}^{2d+1}(g\p_{a_i}g^{-1})+\p_{a_1}(g \delta g^{-1}\prod _{i=2}^{2d+1}(g\p_{a_i}g^{-1}))\big)\nonumber\\
&=&(2d+1)\int{\rm d}\omega{\rm d}^{2d}k\;\ep^{a_1a_2...a_{2d+1}}\tr\big(\p_{a_1}(g \delta g^{-1}\prod _{i=2}^{2d+1}(g\p_{a_i}g^{-1}))\big);
\eea
\end{widetext}
the second-to-last equality follows from the fact that the product term is already a total derivative and with antisymmetricity of the indices, the product of two total derivatives result in another total derivative. The last equality comes from the cyclic property of the trace and the first two terms cancel each other. We therefore have showed that $I_{2d}$ is topological for any $d$.  Notice that even when $g$ satisfies a twisted boundary condition $g(k+G)=U^\dag g(k)U$ on a closed manifold, the variation still vanishes.

In the discussion we have in the main text, however, we have concentrated on non-interacting systems, where the Green's function is taken to be in the form of $(\omega-H_k\pm i\delta)^{-1}.$ We thus are more interested in integrals of Green's functions that are topological subject only to any smooth deformation of $H_k$, instead of an arbitrary deformation of $g$. Fortunately, since  $\p_\omega g^{-1}=1$ with the non-interacting Green's function, we can directly translate the topological combination above to combinations that is invariant under only the deformation of $H_k$. The invariant reads
\beq\label{i2df}
I_F^{2d}=\int{\rm d}\omega{\rm d}^{2d}k\;\ep^{a_1a_2...a_{2d}}\tr\big((\prod _{i=1}^{2d}(g\p_{a_i}g^{-1}))g \big).
\eeq
The indices now only run through all the spatial directions.

Next we show the derivation of Eq.~(\ref{berry}). Since the Berry's phase is invariant under small deformations of the Green's function extended into the surface, the integral has to be topological. By dimensional counting we immediately see that $I^2_F$ is a possible candidate. The rest of the task is to find the constant in front, as well as any possible total derivatives.
\bea
&&\frac12\int_S{\rm d}^2k\epsilon^{ij}\Tr\left(g\p_ig^{-1}\p_j g\right)\nonumber\\
&=&\frac12\int_S{\rm d}^2k\int\frac{{\rm d}\omega}{2\pi}\epsilon^{ij}\tr\left(g\p_ig^{-1}\p_j g\right)\nonumber\\
&=&\frac12\int_S{\rm d}^2k\int\frac{{\rm d}\omega}{2\pi}\epsilon^{ij}\sum_{mn}\frac{\bracket{u_m}{\p_iH_k}{u_n}\braket{\p_ju_n}{u_m}}
{(\omega-E_m)(\omega-E_n)}\nonumber\\
&=&\frac12\int_S{\rm d}^2k\int\frac{{\rm d}\omega}{2\pi}\epsilon^{ij}\sum_{mn}\frac{(E_n-E_m)\braket{u_m}{\p_iu_n}\braket{\p_ju_n}{u_m}}
{(\omega-E_m)(\omega-E_n)}\nonumber\\
&=&\int_S{\rm d}^2k\epsilon^{ij}\sum_{m\in {\rm emp}, n\in {\rm occ}}i\braket{\p_iu_n}{u_m}\braket{u_m}{\p_j u_n}\nonumber\\
&=&\int_S{\rm d}^2k\epsilon^{ij}\sum_{n\in {\rm occ}}i\braket{\p_iu_n}{\p_ju_n}\nonumber\\
&=&\oint_{\p S}{\rm d}k^\ell\sum_{n\in {\rm occ}}i\braket{u_n}{\p_\ell u_n}.
\eea
As we can see, there are no extra total derivative terms. If we integrate this term on a torus, as we did in Eq.~(\ref{charge}), we get
\bea
\frac12\int_{\rm BZ}{\rm d}^2k\epsilon^{ij}\Tr\left(g\p_ig^{-1}\p_j g\right)&=&\int_{BZ}{\rm d}^2k\epsilon^{ij}\sum_{n\in {\rm occ}}i\braket{\p_iu_n}{\p_ju_n}\nonumber\\
&=&2\pi C_1.
\eea

Similarly, we can relate $I^4_F$ to $C_2$, the second Chern number if integrated on a four dimensional manifold without boundaries. In Ref.\cite{QHZ}, they use dispersionless bands to derive the ratio betwenn $I^{4}_F$ and $C_2$. Here we just quote their result (with an extra factor of $(-i)$ since they are using the imaginary time Green's function):
\bea \label{ratio}
C_2&\equiv&\frac{1}{32\pi^2}\int {\rm d}^4k\ep^{ijk\ell}\tr\big(\mathcal{F}_{ij}\mathcal{F}_{k\ell}\big)\nonumber\\
&=&\frac{i}{48\pi^2}I^{4}_F,
\eea
with
\beq
\mathcal{F}_{ij}^{nn'}=\p_i\mathcal{A}^{nn'}_j-\p_j\mathcal{A}^{nn'}_i+i[a_i,a_j]^{nn'}
\eeq
is the Berry's phase gauge field strength. Rather remarkably,  the two integral no longer agree with each other when integrated on a manifold with a boundary. In this case, they differ by a globally defined total derivative.

\section*{Appendix B: From the Green's function expression of the OMP to the expression in energy eigenbasis}
Before we start to evaluate equation (\ref{omp2}), we emphasize again that our distinction between $\alpha_{\rm topo}$ and $\alpha_{\rm exact}$ is different from the distinction between $\alpha_{\rm CS}$ and $\alpha_{\rm G}$ in Ref.\cite{AAJD}. Specifically, $\alpha_{\rm exact}$ is traceless, as can be seen from  Eq.~(\ref{26}), whereas $\alpha_G$ in general is not. With this in mind, let us start from $\alpha_{\rm exact}$:

\beq
{\alpha_{\rm 3d}}_{ij}=-\frac{i}6\ep_{abj}\big((i ab)-c.c.\big).
\eeq

As done in the main text, we use $(iab)$ in short for $\Tr(g\p_ig^{-1}g\p_ag^{-1}g\p_bg^{-1}g)$. The capital trace again denotes tracing over all the bands and integrating over all the momentum and energy divided by $(2\pi)$ for each direction. Expanding in the energy eigenbasis, we get
\begin{widetext}
\bea\label{expand}
(iab)&=&\int\frac{{\rm d}^3k}{(2\pi)^3}\frac{{\rm d}\omega}{2\pi}\sum_{n,\ell, m}\frac{
\bracket{m}{\p_iH}{n}\bracket{n}{\p_aH}{\ell}\bracket{\ell}{\p_bH}{m}}
{(\omega-E_m)^2(\omega-E_n)(\omega-E_\ell)}\nonumber\\
&=&i\int\frac{{\rm d}^3k}{(2\pi)^3}\bigg(-{\rm sgn}(E_n)\theta(-E_nE_\ell)\theta(-E_nE_m)\frac{1}{E_n-E_m}\braket{\p_im}{n}\braket{\p_an}{\ell}\bracket{\ell}{\p_bH}{m} \nonumber\\
&-&{\rm sgn}(E_\ell)\theta(-E_\ell E_m)\theta(-E_\ell E_n)\frac{1}{E_\ell-E_m}\bracket{m}{\p_iH}{n}\braket{\p_an}{\ell}\braket{\p_b\ell}{m}\nonumber\\
&-&{\rm sgn}(E_m)\theta(-E_nE_m)\theta(-E_\ell E_m)\frac{1}{E_n-E_m}\braket{\p_im}{n}\bracket{n}{\p_aH}{\ell}\braket{\p_b\ell}{m}\nonumber\\
&-&{\rm sgn}(E_m)\theta(-E_nE_m)\theta(-E_\ell E_m)\frac{1}{E_\ell-E_m}\braket{\p_im}{n}\bracket{n}{\p_aH}{\ell}\braket{\p_b\ell}{m}\bigg)\nonumber\\
&=& i\int\frac{{\rm d}^3k}{(2\pi)^3}\sum_{m,m'\in {\rm emp}; n,n'\in{\rm occ}}\bigg(\frac{1}{E_n-E_m}
\braket{\p_im}{n}\braket{\p_an}{m'}\bracket{m'}{\p_bH}{m}\nonumber\\
&+&\frac{1}{E_{n'}-E_m}\bracket{m}{\p_iH}{m'}\braket{\p_am'}{n'}\braket{\p_bn'}{m}
-\frac{1}{E_{n'}-E_m}
\braket{\p_im}{n}\bracket{n}{\p_aH}{n'}\braket{\p_bn'}{m}\nonumber\\
&-&\frac{1}{E_n-E_m}\braket{\p_im}{n}\bracket{n}{\p_a H}{n'}\braket{\p_bn'}{m}\bigg)-({\rm occ}\leftrightarrow{\rm emp});\label{mess1}
\eea
we have used $\ket{n}$ to denote $\ket{u_{nk}}$ to avoid cluttering the expression. 

Since $\alpha_{\rm wzw}$ is isotropic, the off-diagonal tensor element comes entirely from $\alpha_{\rm 3d}$. Without loss of generality, let us look at $\alpha_{yz}$:
\bea
\alpha_{yz}&=&-\frac{i}6\ep^{abz}\big((y ab)-c.c.\big)\nonumber\\
&=&-\frac{i}6\bigg(\big((y xy)-(yyx)\big)-c.c.\bigg). 
\eea
In the last equality of Eq.~(\ref{mess1}), the first and the last term are already in the form as derived in Ref.~\cite{AAJD}. Let us look at the two terms in the middle as we first plug in $i=y$, $a=x$, $b=y$:
\bea
&&\sum_{m,m'\in {\rm emp}; n,n'\in{\rm occ}}\bigg(\frac{1}{E_{n'}-E_m}\bracket{m}{\p_yH}{m'}\braket{\p_xm'}{n'}\braket{\p_yn'}{m}
-({\rm occ}\leftrightarrow{\rm emp})\bigg)\nonumber\\
&=&\sum_{m,m'\in {\rm emp}; n,n'\in{\rm occ}}\bigg(\frac{1}{E_{n'}-E_m}\braket{\p_yn'}{m}\bracket{m}{\p_yH}{m'}\braket{\p_xm'}{n'}\nonumber\\
&+&\frac{1}{E_n-E_m}\braket{\p_ym}{n}\bracket{n}{\p_yH}{n'}\braket{\p_xn'}{m}\bigg);
\eea
\bea
&&\sum_{m,m'\in {\rm emp}; n,n'\in{\rm occ}}\bigg(\frac{-1}{E_{n'}-E_m}
\braket{\p_ym}{n}\bracket{n}{\p_xH}{n'}\braket{\p_yn'}{m}-({\rm occ}\leftrightarrow{\rm emp})\bigg)\nonumber\\
&=&\sum_{m,m'\in {\rm emp}; n,n'\in{\rm occ}}\bigg(-\frac{1}{E_{n'}-E_m}\braket{\p_yn'}{m}\braket{\p_ym}{n}\bracket{n}{\p_xH}{n'}\nonumber\\
&-&\frac{1}{E_{n'}-E_m}\braket{\p_ym}{n'}\braket{\p_yn'}{m'}\bracket{m'}{\p_xH}{m}\bigg).
\eea
Similarly, when we plug in $i=y$, $a=y$, $b=x$ and sum with its complex conjugate, we find that every term cancels out:
\bea
\sum_{m,m'\in {\rm emp}; n,n'\in{\rm occ}}\bigg(&&\frac{1}{E_{n'}-E_m}\bracket{m}{\p_yH}{m'}\braket{\p_ym'}{n'}\braket{\p_xn'}{m}
-\frac{1}{E_{n'}-E_m}
\braket{\p_ym}{n}\bracket{n}{\p_yH}{n'}\braket{\p_xn'}{m}\nonumber\\
&-&({\rm occ}\leftrightarrow{\rm emp})\bigg)+c.c.=0.\nonumber\\
\eea
Summing over, we then have
\bea
\bigg(&\big(&(yxy)-(yyx)\big)-c.c.\bigg)\nonumber\\
&=&3i\int\frac{{\rm d}^3k}{(2\pi)^3}\sum_{m,m'\in {\rm emp}; n,n'\in{\rm occ}}\bigg(
\frac{\ep^{ab}}{E_n-E_m}
\braket{\p_ym}{n}\braket{\p_an}{m'}\bracket{m'}{\p_bH}{m}-({\rm occ}\leftrightarrow{\rm emp})
\bigg)-c.c.;
\eea
here $a$ and $b$ run through only $x$ and $y$. We finally have
\beq
\alpha_{yz}=\frac{1}{2}\int\frac{{\rm d}^3k}{(2\pi)^3}\sum_{m,m'\in {\rm emp}; n,n'\in{\rm occ}}\bigg(
\frac{\ep^{ab}}{E_n-E_m}
\braket{\p_ym}{n}\braket{\p_an}{m'}\bracket{m'}{\p_bH}{m}-({\rm occ}\leftrightarrow{\rm emp})
\bigg)+c.c.
\eeq
which is identical to the expression in Ref.~\cite{AAJD}.
\end{widetext}

As for the diagonal components, we have to expand the isotropic term $\alpha_{\rm wzw}$. The calculation is similar to Eq.~(\ref{expand}), except that we have to rearrange terms into total derivatives and then integrate back to the physical momentum space. We currently are not aware of any special trick to automatically rearrange the terms into total derivatives other than the $\mathcal{F\wedge F}$ part, which corresponds to $\alpha_{\rm CS}$ in Ref.~\cite{AAJD}; nevertheless, we can use the Stokes theorem to convert the difference between $\alpha_G$ in Ref.~\cite{AAJD} and $\alpha_{\rm 3d}$ to the extended space and verify that it agrees with the remaining parts in $\alpha_{\rm wzw}$:
\bea\label{diff}
&&(\alpha_G-\alpha_{\rm 3d})_{ij}\nonumber\\
&=&-\frac16\delta_{ij}\int\frac{{\rm d}^3k}{(2\pi)^3}
\ep^{abc}\bracket{m}{\p_aH}{m'}\braket{\p_bm'}{n}\frac{\braket{\p_cn}{m}}{E_n-E_m}+c.c.\nonumber\\
&&-({\rm occ}\leftrightarrow{\rm emp})\nonumber\\
&=&-\frac\pi3\delta_{ij}\int\frac{{\rm d}^4k}{(2\pi)^4}
\ep^{abcd}\p_a\big(\bracket{m}{\p_bH}{m'}\braket{\p_cm'}{n}\frac{\braket{\p_dn}{m}}{E_n-E_m}\big)\nonumber\\
&&+c.c.-({\rm occ}\leftrightarrow{\rm emp}).
\eea
We now start to expand $\alpha_{\rm wzw}$:

\begin{widetext}
\bea
\ep^{abcd}&&\Tr^S({\bf g}\p_a{\bf g}^{-1}{\bf g}\p_b{\bf g}^{-1}{\bf g}\p_c{\bf g}^{-1}{\bf g}\p_d{\bf g}^{-1}{\bf g})
=\int_S\frac{{\rm d}^4k}{(2\pi)^4}\int\frac{{\rm d}\omega}{2\pi}\sum_{ijkl}\ep^{abcd}\frac{\bracket{i}{\p_aH}{j}\bracket{j}
{\p_bH}{k}\bracket{k}{\p_cH}{\ell}\bracket{\ell}{\p_dH}{i}}
{(\omega-E_i)^2(\omega-E_j)(\omega-E_k)(\omega-E_\ell)}\nonumber\\
&\equiv&\int_S\frac{{\rm d}^4k}{(2\pi)^4}
\big(J_{(ik,j\ell)}+(J_{(ij,k\ell)}-c.c.)+J_{(i,jk\ell)}+J_{(k,ij\ell)}
+(J_{(j,ik\ell)}-c.c.)\big))-({\rm occ}\leftrightarrow{\rm emp}).\nonumber\\
\eea
We separate the contributions by different pole placements. $(ik,j\ell)$ for example implies that $E_i$ and $E_j$ are of the same sign and are opposite to $E_k$ and $E_\ell$. Here we illustrate the calculation by explicitly computing the contribution $J_{(ij,k\ell)}$; the calculation for the other contributions are similar and we shall just list the answer. To avoid cluttering the expression we use $i$ to denote $E_i$ as well when there is no ambiguity.
\bea
-iJ_{(ik,j\ell)}&=&\sum_{(ij,kl)}{\rm sgn}(j)\bigg(\frac{(i-j)(j-k)(k-\ell)(\ell-i)}{(j-i)^2(j-k)(j-\ell)}+
\frac{(i-j)(j-k)(k-\ell)(\ell-i)}{(\ell-i)^2(\ell-j)(\ell-k)}\bigg)\ep^{abcd}
\braket{\p_ai}{j}\braket{\p_bj}{k}\braket{\p_ck}{\ell}\braket{\p_d\ell}{i}\nonumber\\
&=&\sum_{(ij,kl)}-{\rm sgn}(j)\bigg(\frac{(k-\ell)(\ell-i)}{(j-i)(j-\ell)}+
\frac{(i-j)(j-k)}{(\ell-i)(\ell-j)}\bigg)\ep^{abcd}
\braket{\p_ai}{j}\braket{\p_bj}{k}\braket{\p_ck}{\ell}\braket{\p_d\ell}{i}\nonumber\\
&=&\sum_{(ij,kl)}{\rm sgn}(j)\bigg(\frac{k-\ell}{j-i}-\frac{k-\ell}{j-\ell}+(j\leftrightarrow\ell)\bigg)\ep^{abcd}
\braket{\p_ai}{j}\braket{\p_bj}{k}\braket{\p_ck}{\ell}\braket{\p_d\ell}{i}\nonumber\\
&=&\sum_{(ij,kl)}{\rm sgn}(j)\bigg(\frac{k-\ell}{j-i}-\frac12\bigg)\ep^{abcd}
\braket{\p_ai}{j}\braket{\p_bj}{k}\braket{\p_ck}{\ell}\braket{\p_d\ell}{i}+c.c.\nonumber\\
&=&\sum_{(ij,kl)}{\rm sgn}(j)\bigg(\frac{k-\ell+j-i}{j-i}-\frac32\bigg)\ep^{abcd}
\braket{\p_ai}{j}\braket{\p_bj}{k}\braket{\p_ck}{\ell}\braket{\p_d\ell}{i}+c.c.\nonumber\\
&=&\sum_{(ij,kl)}{\rm sgn}(j)\bigg(\frac{k-i}{j-i}+\frac{j-\ell}{j-i}-\frac32\bigg)\ep^{abcd}
\braket{\p_ai}{j}\braket{\p_bj}{k}\braket{\p_ck}{\ell}\braket{\p_d\ell}{i}+c.c.\nonumber\\
&=&\sum_{m,m'\in{\rm emp},n,n'\in{\rm occ}}\bigg(2\big(\frac{E_{n'}-E_n}{E_m-E_n}\big)-\frac32\bigg)\ep^{abcd}
\braket{\p_an}{m}\braket{\p_bm}{n'}\braket{\p_cn'}{m'}\braket{\p_dm'}{n}+c.c.\nonumber\\
&-&({\rm occ}\leftrightarrow{\rm emp}).
\eea
\end{widetext}
To get to the first line, we have used 
\beq
\bracket{i}{\p_aH}{j}=\p_aE_i\delta_{ij}+(E_i-E_j)\braket{\p_ai}{j}.
\eeq
In $J_{(ik,j\ell)}$ the first term on the right-hand-side does not contribute. In other parts however that term is important. In the fourth equality, we have taken advantage of the fact that exchanging $j$ and $\ell$ gives the complex conjugate of the whole expression. We symmetrize the second fraction in the third equality with its complex conjugate to get $\frac12$. In the second-to-last equality, the two fractions become each other with a negative sign when we exchange the occupied and empty states, therefore reaching the last equality.

The other contributions are given in the following:
\begin{widetext}
\bea
-i(J_{(ij,k\ell)}-c.c.)&=&\sum_{n,n'\in{\rm occ},m,m'\in{\rm emp}}2\bigg(\frac{E_m-E_m'}{E_n-E_m}+\frac{E_m'-E_m}{E_n-E_m'}\bigg)
\ep^{abcd}\braket{\p_an}{n'}\braket{\p_bn'}{m}\braket{\p_cm}{m'}\braket{\p_dm'}{n}+c.c.\nonumber\\
&+&\sum_{n\in{\rm occ};m,m'\in{\rm emp}}2\bigg(\frac{(E_{m'}-E_{m})\p_aE_n}{(E_m-E_n)^2}\bigg)\ep^{abcd}
\braket{\p_bm}{m'}\braket{\p_cm'}{n}\braket{\p_d n}{m}+c.c.\nonumber\\
&-&({\rm occ}\leftrightarrow{\rm emp});\\
-iJ_{(k,ij\ell)}&=&\sum_{n,n',n''\in{\rm occ};m\in{\rm emp}}\bigg(\frac{(E_n-E_{n'})(E_n-E_{n''})}{(E_n-E_m)^2}\bigg)
\ep^{abcd}\braket{\p_an}{n'}\braket{\p_bn'}{m}\braket{\p_cm}{m'}\braket{\p_dm'}{n}\nonumber\\
&+&\sum_{n\in{\rm occ};m,m'\in{\rm emp}}\bigg(\frac{(E_{m'}-E_m)\p_aE_m}{(E_m-E_n)^2}\bigg)\ep^{abcd}
\braket{\p_bm}{m'}\braket{\p_cm'}{n}\braket{\p_dn}{m}+c.c.\nonumber\\
&-&({\rm occ}\leftrightarrow{\rm emp});\\
-iJ_{(i,jk\ell)}&=&\sum_{n,n',n''\in{\rm occ};m\in{\rm emp}}-\bigg(\frac{(E_n-E_{n'})(E_n-E_{n''})}{(E_n-E_m)^2}\bigg)
\ep^{abcd}\braket{\p_an}{n'}\braket{\p_bn'}{m}\braket{\p_cm}{m'}\braket{\p_dm'}{n}\nonumber\\
&+&\sum_{n,n',n''\in{\rm occ},m\in{\rm emp}}\bigg(\frac{E_{n'}-E_{n''}}{E_m-E_n}-\frac{E_{n'}-E_{n''}}{E_m-E_{n'}}\bigg)
\ep^{abcd}\braket{\p_am}{n}\braket{\p_bn}{n'}\braket{\p_cn'}{n''}\braket{\p_dn''}{m}+c.c.\nonumber\\
&-&\sum_{n\in{\rm occ};m,m'\in{\rm emp}}2\bigg(\frac{(E_{m'}-E_m)\p_aE_m}{(E_m-E_n)^2}\bigg)\ep^{abcd}
\braket{\p_bm}{m'}\braket{\p_cm'}{n}\braket{\p_dn}{m}+c.c.\nonumber\\
&+&\sum_{n\in{\rm occ};m,m'\in{\rm emp}}\bigg(\frac{\p_a(E_{m'}-E_m)}{E_m-E_n}\bigg)\ep^{abcd}
\braket{\p_bm}{m'}\braket{\p_cm'}{n}\braket{\p_dn}{m}+c.c.\nonumber\\
&-&({\rm occ}\leftrightarrow{\rm emp});\\
-i(J_{(j,ik\ell)}-c.c.)&=&\sum_{n,n',n''\in{\rm occ},m\in{\rm emp}}\bigg(\frac{E_{n'}-E_{n''}}{E_m-E_n}-\frac{E_{n'}-E_{n''}}{E_m-E_{n'}}\bigg)
\ep^{abcd}\braket{\p_am}{n}\braket{\p_bn}{n'}\braket{\p_cn'}{n''}\braket{\p_dn''}{m}+c.c.\nonumber\\
&+&\sum_{n\in{\rm occ};m,m'\in{\rm emp}}\bigg(\frac{\p_a(E_{m'}-E_m)}{E_m-E_n}\bigg)\ep^{abcd}
\braket{\p_bm}{m'}\braket{\p_cm'}{n}\braket{\p_dn}{m}+c.c.\nonumber\\
&-&\sum_{n\in{\rm occ};m,m'\in{\rm emp}}\bigg(\frac{(E_{m'}-E_m)\p_aE_m}{(E_m-E_n)^2}\bigg)\ep^{abcd}
\braket{\p_bm}{m'}\braket{\p_cm'}{n}\braket{\p_dn}{m}+c.c.\nonumber\\
&-&({\rm occ}\leftrightarrow{\rm emp});
\eea
\end{widetext}
notice that every term (that is not canceled) come with its complex conjugate. The term in $J_{(ik,j\ell)}$ without energy dependence is the $\mathcal{F\wedge F}$ term since 
\beq
\mathcal F_{ab}^{nn'}=\sum_{m\in emp}-i\braket{\p_an}{m}\braket{\p_bm}{n'}-(a\leftrightarrow b).
\eeq

The remaining terms sum up to be the following total derivative, as one can verify by taking the derivative on every term in the bracket in the following:
\beq
-iJ_{\rm tot}=2\ep^{abcd}\p_a\bigg(\bracket{m}{\p_bH}{m'}\braket{\p_cm'}{n}\frac{\braket{\p_dn}{m}}{E_n-E_m}\bigg).
\eeq
By comparing this expression to Eq.~(\ref{diff}), we finally recover Eq.~(\ref{omp2}).



\begin{thebibliography}{99}
\bibitem{WXZ}Zhong Wang, Xiao-Liang Qi, and Shou-Cheng Zhang, Phys. Rev. Lett. {\bf 105}, 256803 (2010).
\bibitem{king} R. D. King-Smith, and D. Vanderbilt, Phys. Rev. B {\bf 47} 1651 (1993).
\bibitem{niusemi} D. Xiao, J. Shi and Q. Niu, Phys. Rev. Lett. {\bf 95} 137204 (2005).
\bibitem{Thonhauser} T. Thonhauser, D. Ceresoli, D. Vanderbilt, and R. Resta, Phys. Rev. Lett. {\bf 95} 137205 (2005).
\bibitem{ceresoli}D. Ceresoli, T. Thonhauser, D. Vanderbilt, and R. Resta, Phys. Rev. B {\bf 74}, 024408 (2006).
\bibitem{niuq} Junren Shi, G. Vignale, Di Xiao, and Qian Niu, 
Phys. Rev. Lett. {\bf 99}, 197202 (2007).
\bibitem{souza} I. Souza and D. Vanderbilt, Phys. Rev. B {\bf 77}, 054438 (2008).
\bibitem{hasankane} see M. Z. Hasan and C. L. Kane, Rev. Mod. Phys. {\bf 82}, 3045 (2010) or X.-L. Qi and S.-C. Zhang, arxiv:1008.2026
\bibitem{QHZ}Xiao-Liang Qi, Taylor L. Hughes, and Shou-Cheng Zhang, Phys. Rev. B {\bf 78}, 195424 (2008). 
\bibitem{Mala} Malashevich et. al., New J. Phys. {\bf 12}:05032 (2010).
\bibitem{AAJD} A. M. Essin, A.M. Turner, J. E. Moore, and D. Vanderbilt, Phys. Rev. B {\bf 81}, 205104 (2010).
\bibitem{thouless} D. J. Thouless, J. Phys. C {\bf 17}, L325 (1984).
\bibitem{Finkel'stein}M. A. Khodas and A. M. Finkel'stein, Phys. Rev. B {\bf 68}, 155114 (2003) .
\bibitem{zak} J. Zak, Phys. Rev. Lett. {\bf 62}, 2747 (1989).
\bibitem{mywork} Kuang-Ting Chen and Patrick A. Lee, Phys. Rev. B {\bf 83}, 125119 (2011).
\bibitem{mynote} In Ref.~\cite{mywork} we extend to the space where the Brillouin zone becomes the sole boundary. However, this procedure is problematic, when the magnetic field is not in the same direction of the electric field. The magnetic field, while uniform in the physical Brillouin zone, is no longer uniform in the extended space. The same problem occurs when one calculates the higher order corrections in the electric field. By extending the space to a ``cylinder" instead of a ``disk'', one circumvents this problem.
\bibitem{witten}E. Witten, Nucl. Phys. {\bf B 223}, 422 (1983).
\bibitem{gonze} X. Gonze and J. W. Zwanziger, Phys. Rev. B {\bf 84}, 064445, (2011).
 

\end{thebibliography}

\end{document}